\documentclass[prd,aps,nofootinbib,preprintnumbers]{revtex4}
\usepackage[T2A]{fontenc}
\usepackage{amsmath,amssymb}
\usepackage{eucal}
\usepackage{color}
\usepackage[dviwindo]{graphicx}
\newcommand{\bs}{\begin{subequations}}
\newcommand{\es}{\end{subequations}}
\numberwithin{equation}{section}
\newcommand{\ben}{\begin{eqnarray}}
\newcommand{\een}{\end{eqnarray}}
\newcommand{\la}{\label}

\begin{document}

\title{Novel representation of the general Heun's functions}

\author{P. P.  Fiziev
\footnote{fizev@phys.uni-sofia.bg\,\,\,and\,\,\,
fizev@theor.jinr.ru}}
\affiliation{Sofia University Foundation for Theoretical and Computational Physics and Astrophysics, Boulevard
5 James Bourchier, Sofia 1164, Bulgaria\\
and\\
BLTF, JINR, Dubna, 141980 Moscow Region, Rusia}

 \begin{abstract}
In the present article we introduce and study a novel type of solutions of the general Heun's equation.
Our approach is based on the symmetric form of the Heun's differential equation
yielded by development of the Felix Klein symmetric form of the Fuchsian equations
with an arbitrary number $N\geq 4$ of regular singular points.
We derive the symmetry group of these equations which turns to be a proper extension of the Mobius group.
We also introduce and study new series solution of symmetric form of
the general Heun's differential equation ($N=4$)
which treats simultaneously and on an equal footing all singular points.
Hopefully, this new form will simplify the resolution of the existing open problems
in the theory of general Heun's functions and
can be used for development of new effective computational methods.

\vskip .2truecm
PACS numbers: 02.30.Gp, 02.30.Hq
\vskip .2truecm
MSC classification scheme numbers: 34A25, 34B30, 11B37
\end{abstract}
\sloppy
\maketitle

\section{Introduction}
As a tool of the 21st century for solving theoretical, practical and mathematical  problems in all scientific areas,
the Heun's functions are a universal method for treatment of a vast variety of phenomena
in complicated systems of different kinds:
in solid state physics, crystalline materials, graphene, in celestial mechanics, quantum mechanics,
quantum optics, quantum field theory, atomic and nuclear physics, heavy ion physics,
hydrodynamics, atmosphere physics, gravitational physics, black holes, compact stars,
and especially in extremely urgent and expensive search for gravitational waves,
astrophysics, cosmology, biophysics, studies of the genome structure, mathematical chemistry,
economic and financial problems, etc.
This wide area of application is a result of the general type of the Heun's
differential equation that properly describes processes in all scientific areas.

The general Heun's equation written in the Fuchsian form
\ben
\hskip -.truecm
H''+\left({\frac {\gamma_{{}_G}} {z}}+{\frac{\delta_{{}_G}}{z-1}}+{\frac{\epsilon_{{}_G}}{z-a_{{}_G}}}\right)H'+
{\frac{\alpha_{{}_G}\beta_{{}_G} z-\lambda}{z(z-1)(z-a_{{}_G})}}H = 0,\,\,\,\,\,
\gamma_{{}_G}\!+\!\delta_{{}_G}\!+\!\epsilon_{{}_G}\!=\!\alpha_{{}_G}\!+\!\beta_{{}_G}\!+\!1;
\hskip 1truecm
\la{dHeunG}
\een
was constructed by Karl Heun in  \cite{Heun} as a generalization of the standard hypergeometric equation
by adding one more regular singular point in complex plane: $z=a_{{}_G}\in \mathbb{C}$.
\footnote{Everywhere in this paper the prime denotes a derivative with respect to the variable $z$.}

At present, this is the most popular form of the Heun's equation, see \cite{Ron,SL} and the literature therein.
It is not symmetric with respect to four regular singular points $0, 1, a_{{}_G}, \infty$ with the corresponding indices
\ben
\{ 0, 1-\gamma_{{}_G}\},\quad \{ 0, 1-\delta_{{}_G} \},\quad \{ 0, \gamma_{{}_G}+\delta_{{}_G}-\alpha_{{}_G}-\beta_{{}_G} \},\quad \{ \alpha_{{}_G}, \beta_{{}_G} \}.
\la{indexG}
\een

The Heun's general function
$\text{HeunG}(a_{{}_G}, \lambda, \alpha_{{}_G}, \beta_{{}_G}, \gamma_{{}_G}, \delta_{{}_G}, z)$
is defined as the unique local regular solution around the regular singular point $z=0$ under normalization
$\text{HeunG}(a_{{}_G}, \lambda, \alpha_{{}_G}, \beta_{{}_G}, \gamma_{{}_G}, \delta_{{}_G}, 0)=1$
\footnote{Here we are using the notations of the widespread computer package Maple.}.
The second linearly independent local solution can be obtained via a proper change of the  parameters,
as described, for example, in \cite{Ron,SL}. Using proper Mobius transformations (See the Appendix \ref{ApA}.) one can also obtain
similar local solutions around  other regular singular points implementing the Heun's general function
(see, for example, \cite{Ron,SL}). Thus, the problem of finding all local solutions of Eq. \eqref{dHeunG}
is reduced to the study of the Heun's general function defined in the vicinity of the point $z=0$ by the absolutely
convergent series
\ben
\text{HeunG}(a_{{}_G}, \lambda, \alpha_{{}_G}, \beta_{{}_G}, \gamma_{{}_G}, \delta_{{}_G},z)=
\sum_{n=0}^\infty h_{n}(a_{{}_G}, \lambda, \alpha_{{}_G}, \beta_{{}_G}, \gamma_{{}_G}, \delta_{{}_G})z^n.
\la{HeunG}
\een
Replacing the function $H(z)$ in Eq. \eqref{dHeunG} with the series \eqref{HeunG}
one easily obtains the simple three-term recurrence relation
\ben
h_{n}+R_{n-1}h_{n-1}+R_{n-2}h_{n-2}=0
\la{recurrenceG}
\een
with the coefficients
\ben
R_{n-1}= -1 -{\frac 1 {a_{{}_G}}} +
{\frac {\lambda-\gamma_{{}_G}(a_{{}_G}\delta_{{}_G}-a_{{}_G}+\alpha_{{}_G}+\beta_{{}_G}-\delta_{{}_G}-\gamma_{{}_G})} {a_{{}_G}(\gamma_{{}_G} +n-1)(\gamma_{{}_G}-1)}},\nonumber \\
R_{n-2}={\frac 1 {a_{{}_G}}} + {\frac {-\alpha_{{}_G}\beta_{{}_G}+\alpha_{{}_G}\gamma_{{}_G}+\beta_{{}_G}\gamma_{{}_G}-\gamma_{{}_G}^2+\alpha_{{}_G}+\beta_{{}_G}-2\gamma_{{}_G}-1} {a_{{}_G}(\gamma_{{}_G} +n-1)(\gamma_{{}_G}-1)}}.
\la{R}
\een

Using relations \eqref{recurrenceG}, \eqref{R} and the initial conditions $h_0=1, h_1=\lambda/a_{{}_G}\gamma_{{}_G}$
one can effectively calculate the values of the series \eqref{HeunG} in the circle around the point $z=0$ with the circle-radius $<1$,
i.e., before approaching the next regular singular point $z=1$.

Trying to continue the series \eqref{HeunG} outside this circle, one meets hard numerical problems,
as seen from the ten-year not very satisfactory attempts to improve the only existing computer code for work
with the Heun's functions -- Maple.
At present, this is a serious obstacle for numerous applications of these extremely useful functions.

The main idea of the present paper is to find a novel representation of the solutions of the general Heun's equation
which gives an equal treatment of all regular singular points and yields series expansions
which are valid simultaneously in the vicinities of all of them.

We succeeded in finding such an approach but, as one can expect,
it leads to new and more complicated series expansions
of solutions of the general Heun's functions defined by the nine-term recurrence relations.
Fortunately, such recurrence relations are not a problem for modern computers.
In the present paper, we introduce for the first time these new series and study their basic properties.

One can hope that the new series will be a
useful tool for solution of the basic open problems in the theory of the general Heun's functions, like
connection problem, study of the asymptotics, monodromy group, relations between the derivatives of the Heun's functions,
e.t.c.,
as well as for development of new more efficient computational techniques.

\section{Symmetric form of the Heun's equation}

\subsection{The symmetric form of the general Fuchsian equation}

The {\em symmetric} form of the general Fuchsian equation with $N\geq 4$ arbitrary
regular singular points $z_{j=1,...,N}\in \mathbb{C}$ was adopted
by Felix Klein as early as in \cite{Klein}:
\ben
\mathcal{W}''+\left(\sum_{j=1}^N{\frac{1-\alpha_j-\beta_j}{z-z_j}}\right)\mathcal{W}'+
{\frac 1 {P(z)}\left(\Lambda(z)+\sum_{j=1}^N{\frac{q_j}{z-z_j}}\right)}\mathcal{W} = 0,
\la{dWN}
\een
see also \cite{Forsyth}. Here
\ben
\hskip -1.truecm P(z)=\prod_{j=1}^N(z-z_j)=\sum_{n=0}^N(-1)^n\sigma_{{}_{N-n}} z^n,\quad \Lambda(z)=\sum_{l=0}^{N-4}\lambda_l z^l,\quad
\text{and}\,\,\,q_j=\alpha_j\beta_j P^\prime(z_j)\,\,\, \text{for}\,\,\,{j=1,...,N}.
\la{LP}
\een

Under the additional condition
\ben
\sum_{j=1}^N \left(\alpha_j+\beta_j\right)=N-2
\la{constr}
\een
the point $z=\infty$ is a regular one for Eq. \eqref{dWN}. Thus,
it remains with only $N$ finite regular singular points $z_{j=1,...,N}\in \mathbb{C}$
with arbitrary indices $\{\alpha_j,\beta_j\}_{j=1...N}\in \mathbb{C}$.
As seen, such equations are determined altogether by $4(N-1)$ arbitrary complex numbers:
singular points $z_{j=0,...,N}$, their indices $\{\alpha_j,\beta_j\}:{j=0,...,N}$ with constraint \eqref{constr},
and auxiliary parameters $\lambda_{l=0,...,N-4}\in \mathbb{C}$.

Now one can use the following transformation of the unknown function $\mathcal{W}(z)$ with the properly chosen parameters $\nu_{j=1,...,N}$:
\ben
\mathcal{W}(z) = \mathcal{F}(z)\prod_{j=1}^N \left(z-z_j\right)^{\nu_j}, \qquad \sum_{j=1}^N \nu_j=0,
\la{nu_transf}
\een
to fix $N-1$ of the parameters $\{\alpha_j,\beta_j\}:{j=0,...,N}$, or some $(N-1)$-in-number their combinations.
The second condition in Eq. \eqref{nu_transf} is necessary to preserve  relation \eqref{constr}.
As a result of the last two constraints, we remain with altogether $N$ free parameters between the indices $\{\alpha_j,\beta_j\}:{j=0,...,N}$.
For example, an asymmetric choice $\beta_{j=1,...,N-1}=0$, similar to \eqref{indexG}, is possible.
Thus, we remain with altogether $3(N-1)$ free complex parameters.

Instead of the above asymmetric choice, which destroys the symmetric treatment of the regular singular points,
we prefer to use the following $N$-in-number {\it symmetric} constraints on the indices $\{\alpha_j,\beta_j\}:{j=0,...,N}$:
\ben
\alpha_j+\beta_j=1 - {\tfrac 2 N}, \quad \alpha_j\beta_j P^\prime(z_j)=q_j,\quad \text{for}\quad j=1,...,N,
\la{index}
\een
thus preserving equal treatment of all $N$ regular singular points and relation \eqref{constr}.
Introducing new $N$-in-number free uniformization parameters $\chi_{j=1,...,N} \in \mathbb{C}$
we obtain for all $j=1,...,N$:
\ben
\alpha_j=\left(1-{\tfrac 2 N}\right) \left(\cos\chi_j\right)^2,\quad \beta_j=\left(1-{\tfrac 2 N}\right) \left(\sin\chi_j\right)^2,\nonumber\\
\quad q_j=\left(\left({\tfrac 1 2}-{\tfrac 1 N}\right)\sin(2\chi_j)\right)^2 \prod_{k\neq j}^N\left(z_j-z_k\right).
\la{alphabeta}
\een
and Eq. \eqref{dWN} acquires its simplest $3(N-1)$-parameter final form:
\ben
\mathcal{F}''+{\frac 2 N}\left(\sum_{j=1}^N{\frac 1{z-z_j}}\right)\mathcal{F}'+
{\frac 1 {P(z)}\left(\Lambda(z)+\sum_{j=1}^N{\frac{q_j}{z-z_j}}\right)}\mathcal{F} = 0,
\la{dWFsym}
\een

Note that:

i) In Eq. \eqref{dWFsym} one can consider the parameters $q_{j=1,...,N}$ as independent ones,
instead of the uniformization parameters $\chi_{j=1,...,N}$.
The disadvantage of this approach is in the introduction of branching points
of the indices $\{\alpha_j,\beta_j\}:{j=1,...,N}$,
since indices of singular points of Eq. \eqref{dWFsym} are the roots $x_j^\pm$ of the corresponding quadratic equations
$x_j^2-2\left({\tfrac 1 2}-{\tfrac 1 N}\right)x_j +q_j/P^\prime(z_j)=0$, $j=1,...,N$.
The presence of such branching points is undesirable
since it requires special care during numerical calculations.

ii)  We still have the freedom to lower the number of the free parameters of the problem,
moving some three different singular points
to any convenient different places in the complex plane $\mathbb{C}$,
for example to $0,1,a_{{}_G}$, as in the case of general Heun's functions.
This can be done by using proper Mobius transformation
without changing the number and the character of the singular points of Eq. \eqref{dWFsym},
see Appendix \ref{ApA}.
Thus, we will end with $3(N-2)$ essential free parameters of Eq. \eqref{dWFsym}, as illustrated
in the rest of the paper by the basic example $N=4$
\footnote{For another form of the general Fuchsian equation with $N$ singular points
and the corresponding count of the number of free parameters in it see \cite{NIST}.}.

iii) There exist two quite different cases of positions of the singular points $z_{j=0,...,N}$.

$\bullet$ The first one is the special case when all regular singular points $z_{j=0,...,N}$ of Eq. \eqref{dF} lie
on some circle $\mathfrak{C}\in \mathbb{C}$.
A special and natural case is the one when $z_{j=0,...,N}\in \mathbb{R}$, i.e.,
all singular points are real, as in the important Smirnov's Thesis  \cite{Smirnov}.
We will call this case {\it the circular case}.
In the circular case, one is able to move all singular points $z_{j=0,...,N}$ on any other circle
$\mathfrak{\tilde C}\in \mathbb{C}$ using Mobius transformation, see Appendix \ref{ApA}.

$\bullet$ The opposite (general) case is the one in which the singular points $z_{j=0,...,N\geq 4}$ do not lie
on any circle in the complex plane. We call it  {\it the non circular case}.
In the non circular case, the theory of solutions of Eq. \eqref{dF} is much more complicated.
It is not developed enough even for the general Heun's equation \eqref{dHeunG}.

Hence, the choice of the position of the regular singular points $z_{j=0,...,N}$ of Eq. \eqref{dF}
is an important component of the general theory. In the present paper, we investigate this problem only
for the circular case with $N=4$, since it corresponds to the general Heun's functions.
As we shall show, in this case the general theory is a relatively simple one.

Equation \eqref{dWFsym} can be written down also in the following self-adjoint form:
\ben
\left(P(z)\right)^{1-2/N}\left(P(z)^{2/N}\mathcal{F}'\right)^\prime+
\left(\Lambda(z)-\left({\tfrac 1 2}-{\tfrac 1 N}\right)^2{\tfrac 1 {P(z)}}
\sum_{j=1}^N \left(\sin(2\chi_j)\right)^2\partial_z P(z_j)\partial_{z_j}P(z) \right)\mathcal{F} = 0.
\la{dWFsymSA}
\een

{\bf Proposition 1:} Let $\Lambda_{1,2}(z)$ be two polynomials of degree $(N-4)$ with equidistant coefficients:
$\lambda_{2,l}-\lambda_{1,l}=\Delta\lambda\neq 0\quad \forall\quad l=0,...,(N-4)$. Hence,
\ben
\Lambda_2(z)-\Lambda_1(z)=\Delta\lambda\,{\frac{z^{N-3}-1}{z-1}}.
\la{Lambda_1-2}
\een
Then, the solutions $\mathcal{F}_{\Lambda_{1,2}}(z)$ of the corresponding Eqs. \eqref{dWFsym}, or \eqref{dWFsymSA}
are orthogonal with respect to the measure
\ben
d\mu(z)={\frac{z^{N-3}-1}{z-1}}\left(P(z)\right)^{{\frac 2 N}-1},\quad\text{i.e.,}
\la{measureN}
\een
\ben
\int_{z_j}^{z_j}\mathcal{F}_{\Lambda_{1}}(z)\mathcal{F}_{\Lambda_{2}}(z)d\mu(z)=0,
\la{orthogonalN}
\een
if the boundary conditions
\ben
\left(P(z)\right)^{{\frac 2 N}}
\left(\mathcal{F}_{\Lambda_{1}}(z)\mathcal{F}_{\Lambda_{2}}^\prime(z)-
\mathcal{F}_{\Lambda_{2}}(z)\mathcal{F}_{\Lambda_{1}}^\prime(z)\right)\,\upharpoonleft_{z_j,z_j}=0
\la{boundaryN}
\een
are satisfied.\,$\blacktriangleleft$ \footnote{The
sign $\blacktriangleleft$ denotes the end of the corresponding statement.}

Thus, we arrived at a quite unusual boundary-value problem for Eqs. \eqref{dWFsym} and \eqref{dWFsymSA}.

The proof is based on the integration of the general identity valid for any functions $\Lambda_{1,2}(z)$ in Eqs. \eqref{dWFsym} and \eqref{dWFsymSA}:
\ben
\left(\Lambda_2(z)-\Lambda_1(z)\right)\left(P(z)\right)^{{\frac 2 N}-1}\mathcal{F}_{\Lambda_{1}}(z)\mathcal{F}_{\Lambda_{2}}(z)\equiv
\left(\left(P(z)\right)^{{\frac 2 N}} \left(\mathcal{F}_{\Lambda_{1}}(z)\mathcal{F}_{\Lambda_{2}}^\prime(z)-
\mathcal{F}_{\Lambda_{2}}(z)\mathcal{F}_{\Lambda_{1}}^\prime(z)\right)\right)^\prime.
\la{identityN}
\een

{\bf Proposition 2:}  Equations \eqref{dWFsym} and \eqref{dWFsymSA} are invariant under
the {\it extension} of the Mobius group $\widehat{\mathfrak{G}}_{Mobius}$
that acts on the functions of $(3N-2)$ variables
$\mathcal{F}\left(z;z_1,...,z_N;q_1,...,q_N;\lambda_0,...\lambda_{N-4}\right)$
and is produced by the following basic transformations:

\begin{description}
  \item[(i)] Complex translations with arbitrary $\zeta \in \mathbb{C}$:
\ben
\hskip - 1.2truecm z\to  z+\zeta; \quad z_j\to z_j+\zeta:\, j=1,...,N;\quad q_j\to q_j:\, j=1,...,N; \quad
\lambda_l\to \sum_{m=l}^{N-4} {\binom {m}{l}}\zeta^{m-l}\lambda_{m}:\, l=0,...,N-4.
\la{translationN}
\een
  \item[(ii)] Complex dilatations with arbitrary $t\in \mathbb{C}$, $t\neq 0$:
\ben
z\to t\, z; \quad z_j\to t\, z_j:\, j=1,...,N; \quad q_j\to t^{N-1} q_j:\, j=1,...,N;\quad \lambda_l\to t^{N-l-2} \lambda_l:\, l=0,...,N-4.
\la{rescalingN}
\een
  \item[(iii)] Inversion
\ben
z&\to& 1/ z; \quad z_j\to 1/ z_j:\, j=1,...,N; \quad q_j \to {\tfrac {(-1)^{N-1}}{\sigma_{\!{}_N}}} z_j^{1-N} q_j :\, j=1,...,N;\nonumber\\
\lambda_l &\to& {\tfrac {(-1)^{N-1}}{\sigma_{\!{}_N}}}\bigg(\Big(\sum_{j=1}^N z_j^{l+3-N}q_j\Big)-\lambda_{N-4-l}\bigg):\, l=0,...,N-4;
\la{inversionN}
\een
where $\sigma_{\!{}_N}=\prod_{j=1}^N z_j$.\nonumber\,$\blacktriangleleft$
\end{description}

Indeed, it is not hard to check directly the invariance of Eq. \eqref{dWFsym}
under transformations \eqref{translationN}, \eqref{rescalingN}, and \eqref{inversionN}.

Using proper compositions of these basic transformations (see Appendix \ref{ApA}) we are able to construct
a representation of the whole extended Mobius group $\widehat{\mathfrak{G}}_{Mobius}$ that acts on the solutions
$\mathcal{F}\left(z;z_1,...,z_N;q_1,...,q_N;\lambda_0,...\lambda_{N-4}\right)$ of Eq. \eqref{dWFsym}
without bringing us outside of the variety of these solutions. Hence, $\widehat{\mathfrak{G}}_{Mobius}$
is the group of invariance of the variety of solutions to Eq. \eqref{dWFsym}.

\subsection{Symmetric form of the general Heun's equation and its Taylor series solutions}

\subsubsection{Symmetric form of the general Heun's equation.}
The symmetric form \eqref{dWN} for the special case of the general Heun's Eq. \eqref{dHeunG}
(i.e., for $N=4$) was pointed out in \cite{Ron}.
For brevity, in this case, we denote by $\lambda$ the single auxiliary parameter.
Then, the symmetric form \eqref{dWFsym} of the general Heun's Eq. \eqref{dHeunG} reads
\ben
\mathcal{F}^{\prime\prime}+{\frac 1 2}\left(\sum_{j=1}^4{\frac 1{z-z_j}}\right)\mathcal{F}^\prime+
{\frac 1 {P(z)}\left(\lambda+\sum_{j=1}^4{\frac{q_j}{z-z_j}}\right)}\mathcal{F} = 0,
\la{dF}
\een
or in a self-adjoint form:
\ben
\left(P(z)\right)^{1/2}\left(\left(P(z)\right)^{1/2}\mathcal{F}^\prime\right)^\prime +
\left(\lambda+Q(z)\right)\mathcal{F} = 0,
\la{dFsa}
\een
where $Q(z)=\sum_{j=1}^4{\frac{q_j}{z-z_j}}=-{\tfrac 1 {16}}{\tfrac 1 {P(z)}}
\sum_{j=1}^4 \left(\sin(2\chi_j)\right)^2\partial_z P(z_j)\partial_{z_j}P(z)$.

\subsubsection{The orthogonality of solutions.}
The form \eqref{dFsa} shows that the auxiliary parameter $\lambda$ actually plays the role of eigenvalue of the problem.
This form is also convenient for discussing the orthogonality of the solutions
$\mathcal{F}$ on the contours $\mathcal{L}\in\mathbb{C}$ under the measure $d\mu(z)=\left(P(z)\right)^{-1/2}dz$ \footnote{For polynomial $P(z)$
of the fourth degree this measure is obviously related
with the elliptic integrals \cite{NIST},
thus giving the basis for the well-known relation of the general Heun's functions
with elliptical ones, see, for example,  \cite{Fiziev10} and the references therein.}:

{\bf Proposition 3:}
For any two solutions $\mathcal{F}_{\lambda_{1,2}}(z)$  of the Eq. \eqref{dFsa} with  $\lambda_1\neq\lambda_2$ we have
\ben
\hskip -.5truecm \int_{\mathcal{L}_{ij}}\mathcal{F}_{\lambda_1}(z)\mathcal{F}_{\lambda_2}(z)d\mu(z)=0.
\la{orthogonality}
\een
Here $\mathcal{L}_{ij}\in\mathbb{C}$ is any contour which starts at the singular point $z_j$ and ends at the singular point $z_j$
without going through the other singular points $z_{k\neq i,j}$. Besides, the singular boundary conditions
\ben
\left(P(z)\right)^{1/2}
\left(\mathcal{F}_{\lambda_1}(z)\mathcal{F}_{\lambda_2}^\prime(z)-
\mathcal{F}_{\lambda_2}(z)\mathcal{F}_{\lambda_1}^\prime(z)\right)\,\upharpoonleft_{z_j,z_j}=0
\la{boundary}
\een

\noindent are supposed to be satisfied. The same boundary conditions ensure the self-adjoint property of the differential operator
in Eq. \eqref{dFsa} with respect to the measure $d\mu(z)=\left(P(z)\right)^{-1/2}dz$.\,$\blacktriangleleft$.

Indeed, the orthogonality relation \eqref{orthogonality} is an immediate consequence of the identity
\ben
\hskip -1.3truecm \left(\lambda_2 -\lambda_1\right) \int_{\mathcal{L}_{ij}}\mathcal{F}_{\lambda_2}(z)\mathcal{F}_{\lambda_1}(z)d\mu(z)=
\left(P(z)\right)^{1/2}
\left(\mathcal{F}_{\lambda_2}(z)\mathcal{F}_{\lambda_1}^\prime(z)-
\mathcal{F}_{\lambda_1}(z)\mathcal{F}_{\lambda_2}^\prime(z)\right)\,\Big|_{z_j,z_j},
\la{identity}
\een
which follows from Eq. \eqref{dFsa} by applying the well-known procedure for two solutions $\mathcal{F}_{\lambda_{1,2}}(z)$
with $\lambda_1 \neq\lambda_2$, and yields the boundary conditions \eqref{boundary}.

It is not hard to justify Proposition 3.
Indeed, Eq. \eqref{dF} has the following two linearly independent local Frobenius
solutions in the vicinity of each regular singular point $z_j$ \cite{Golubev, CL}:
\ben
\hskip -.6truecm \mathcal{F}_{\alpha_j}(z)=\left(z-z_j\right)^{\alpha_j}\sum_{n=0}^\infty f_{\alpha_j, n}\left(z-z_j\right)^n,\quad
\mathcal{F}_{\beta_j}(z)=\left(z-z_j\right)^{\beta_j} \sum_{n=0}^\infty f_{\beta_j, n}\left(z-z_j\right)^n.
\la{Frobenius}
\een
Then, in the vicinity of the point $z_j$ the solutions $\mathcal{F}_{\lambda_{1,2}}(z)$ allow the representation
\ben
\mathcal{F}_{\lambda_{1,2}}(z)=C^{\alpha_j}_{\lambda_{1,2}}\mathcal{F}_{\alpha_j}(z)+C^{\beta_j}_{\lambda_{1,2}}\mathcal{F}_{\beta_j}(z)
\la{localF_z_j}
\een
with proper constants $C^{\alpha_j}_{\lambda_{1,2}},C^{\beta_j}_{\lambda_{1,2}}$. Taking into account that in the same vicinity
$P(z)=P^\prime(z_j)(z-z_j)+O_2(z-z_j)$, one obtains from Eq. \eqref{localF_z_j}
\ben
\left(P(z)\right)^{1/2}
\left(\mathcal{F}_{\lambda_2}(z)\mathcal{F}_{\lambda_1}^\prime(z)-
\mathcal{F}_{\lambda_1}(z)\mathcal{F}_{\lambda_2}^\prime(z)\right)=\\
=\left(\alpha_j-\beta_j\right)\left(P^\prime(z_j)\right)^{1/2}
\left(C^{\alpha_j}_{\lambda_{1}}C^{\beta_j}_{\lambda_{2}}-C^{\alpha_j}_{\lambda_{2}}C^{\beta_j}_{\lambda_{1}}\right) +\mathcal{O}(z-z_j).
\nonumber
\la{orth_cond}
\een

Hence, the boundary condition at the singular point $z_j$ can be satisfied either if
$\alpha_j=\beta_j$ which gives $\chi_j/\hskip -.2truecm \mod\!(2\pi)=\pm \pi/4, \pm 3\pi/4$
or if $C^{\alpha_j}_{\lambda_{1}}/C^{\beta_j}_{\lambda_{1}}= C^{\alpha_j}_{\lambda_{2}}/C^{\beta_j}_{\lambda_{2}}$
which leads to a special coherent choice of the solutions $\mathcal{F}_{\lambda_{1,2}}(z)$ \eqref{localF_z_j}
with the coefficient ratio being independent of the eigenvalues $\lambda_{1,2}$.

If one imposes the same boundary condition also at the second singular point $z_j\neq z_j$,
then one arrives at a specific two-singular-point boundary problem \cite{Ron,SL,KLS,LS}.
In this case the auxiliary parameter of the solution, i.e. the eigenvalue $\lambda$, can have only
some definite values which define the spectrum of the self-adjoint operator in Eq. \eqref{dFsa}, see \cite{Smirnov},
where the standard approach to the general Heun's functions was substantially elaborated.
This confirms once again our interpretation of the auxiliary parameter $\lambda$ as an eigenvalue parameter
in Eq. \eqref{dFsa}.

Note that in our approach it is possible to impose simultaneously
regular boundary conditions at two regular singular points
since $\mathcal{F}(z)$ is not the standard {\it local} solution like
$\text{HeunG}(a_{{}_G}, \lambda, \alpha_{{}_G}, \beta_{{}_G}, \gamma_{{}_G}, \delta_{{}_G}, z)$\footnote{We
remind the reader that the term {\it general-Heun's-function} is in use for
the local solution $\text{HeunG}(a_{{}_G}, \lambda, \alpha_{{}_G}, \beta_{{}_G}, \gamma_{{}_G}, \delta_{{}_G}, z)$,
and can not be applied to any other solution, like $\mathcal{F}(z)$, to the general Heun's equation.}.
In the last case, the regularity condition is already imposed at the point $z=0$ by definition.
Therefore, one is able to impose on the local regular solutions like
$\text{HeunG}(a_{{}_G}, \lambda, \alpha_{{}_G}, \beta_{{}_G}, \gamma_{{}_G}, \delta_{{}_G}, z)$
only one more regularity condition at some different regular singular point\footnote{The
author is grateful to Professor S. Yu. Slavyanov for this remark,
as well as for drawing the author's attention to reference \cite{Smirnov}.}.
Obviously, the last (widely accepted) approach is equivalent to ours.
Our treatment seems to be more natural and corresponds to the standard
boundary problem for an ordinary differential equation (See also \cite{Smirnov}.).

\subsubsection{Elementary symmetric functions related to the problem.}
Further on, we use the representation $P(z)=z^4-\sigma_1 z^3+\sigma_2 z^2-\sigma_3 z +\sigma_4$
of this fourth-degree-polynomial, thus introducing the standard elementary symmetric functions
\ben
\sigma_1=z_1+z_2+z_3+z_4, \quad
\sigma_2=z_1z_2+z_1z_3+z_1z_4+z_2z_3+z_2z_4+z_3z_4,\nonumber\\
\sigma_3=z_2z_3z_4+z_1z_3z_4+z_1z_2z_4+z_1z_2z_3,\quad
\sigma_4=z_1z_2z_3z_4,
\la{sigma}
\een

\noindent and the additional notation
\ben
\sigma_1^j=\sigma_1(z_j=0),\quad\text{for example,}\quad \sigma_1^1=z_2+z_3+z_4,\quad \text{etc},\nonumber\\
\sigma_2^j=\sigma_2(z_j=0),\quad\text{for example,}\quad \sigma_2^1=z_2z_3+z_2z_4+z_3z_4,\quad  \text{etc},\\
\sigma_3^j=\sigma_3(z_j=0),\quad\text{for example,}\quad \sigma_3^1=z_2z_3z_4, \quad  \text{etc} .\nonumber
\la{sigma_j}
\een

\subsubsection{Invariance of the symmetric form of the general Heun's equation under inversion.}
Now it is easy to check that the symmetric form of the general Heun's equation Eq. \eqref{dF} (as well as Eq. \eqref{dFsa})
is covariant under a proper extension of the Mobius group $\mathfrak{G}_{Mobius}$. Indeed, applying the results of Proposition 2,
in the case $N=4$ we obtain much simpler results.

{\bf Proposition 4:}

Equation \eqref{dF} is invariant under the {\it extension} of the Mobius group $\widehat{\mathfrak{G}}_{Mobius}$
that acts on the functions of $10$ variables $\mathcal{F}\left(z;z_1,...,z_4;q_1,...,q_4;\lambda\right)$
and is produced by the following basic transformations:

\begin{description}
  \item[(i)] Complex translations with arbitrary $\zeta \in \mathbb{C}$:
\ben
\hskip - 1.2truecm z\to  z+\zeta; \quad z_j\to z_j+\zeta:\, j=1,...,4;\quad q_j\to q_j:\, j=1,...,4; \quad
\lambda \to \lambda.
\la{translationN4}
\een
  \item[(ii)] Complex dilatations with arbitrary $t\in \mathbb{C}$, $t\neq 0$:
\ben
z\to t\, z; \quad z_j\to t\, z_j:\, j=1,...,4; \quad q_j\to t^3 q_j:\, j=1,...,4;\quad \lambda \to t^2 \lambda.
\la{rescalingN4}
\een
  \item[(iii)] Inversion
\ben
\hskip -1.4truecm z&\to& 1/ z; \quad z_j\to 1/ z_j:\, j=1,...,4; \quad q_j \to -q_j/\left( z_j^2 \sigma_4 \right) :\, j=1,...,4;\quad
\lambda \to \Big(\lambda -\sum_{j=1}^4 q_j/z_j\Big)/\sigma_4.\,  \blacktriangleleft
\la{inversionN4}
\een
\end{description}

Using proper compositions of these basic transformations we are able to construct
a representation of the whole extended Mobius group $\widehat{\mathfrak{G}}_{Mobius}$ that acts on the solutions
$\mathcal{F}\left(z;z_1,...,z_4;q_1,...,q_4;\lambda\right)$ of Eq. \eqref{dF}
without bringing us outside of the variety of these solutions. The extended group $\widehat{\mathfrak{G}}_{Mobius}$
is the group of invariance of the variety of solutions to Eq. \eqref{dF}.

\subsubsection{The Taylor series expansion of the solutions.}
Our next step is to adopt the following basic assumption which is of crucial importance for further work:
\ben
z_{j=0,...,4}\neq 0.
\la{zero}
\een
Then the function $\mathcal{F}(z)$
is an analytical one in the vicinity of the point $z=0$ and has an absolutely convergent
Taylor series expansion
\ben
\mathcal{F}(z)\equiv\mathcal{F}(z;z_1,...z_4;q_1,...q_4;\lambda) = \sum_{n=0}^\infty f_n(z_1,...z_4;q_1,...q_4;\lambda) z^n
\la{taylorF}
\een
with the coefficients $f_n(q_1,...,q_4;\lambda)$ defined by the nine-term recurrence relation
\ben
f_n+\sum_{k=1}^8 r_{n-k}f_{n-k}=0,
\la{frecurrence}
\een
which can be obtained from Eqs. \eqref{dF} and \eqref{taylorF}.

After some lengthly but straightforward calculations one derives the following relations
for eight coefficients $r_{n-1},...,r_{n-8}$:
\begin{subequations}\label{r:abcdefgh}
\ben
(\sigma_4)^2r_{n-1}&=&-\left( 2-{\tfrac  7 2}{\tfrac  1 n}\right)\sigma_3\sigma_4,\label{r:a}\\
(\sigma_4)^2r_{n-2}&=&{\tfrac{\sigma_4}{n(n-1)}}\Big(\lambda- \sum_{j=1}^4 q_j/z_j\Big)-
\left(1-{\tfrac  5 n}+{\tfrac  3 2}{\tfrac  1 {n-1}}\right)\left((\sigma_3)^2+2\sigma_2\sigma_4\right),\label{r:b}\\
(\sigma_4)^2r_{n-3}&=&-{\tfrac 1 {n(n-1)}}\Big(\lambda\sigma_3- \sum_{j=1}^4 q_j\sigma_2^i\Big)-
\left(2-{\tfrac  {39} 2}{\tfrac  1 n}+{\tfrac  9 {n-1}}\right)\left(\sigma_2\sigma_3+\sigma_1\sigma_4\right),\label{r:c}\\
(\sigma_4)^2r_{n-4}&=&{\tfrac 1 {n(n-1)}}\Big(\lambda\sigma_2- \sum_{j=1}^4 q_j\sigma_1^i\Big)+
\left(1-{\tfrac  {16} n}+{\tfrac  9 {n-1}}\right)\left((\sigma_2)^2+2\sigma_1\sigma_3+2\sigma_4\right),\label{r:d}\\
(\sigma_4)^2r_{n-5}&=&-{\tfrac 1 {n(n-1)}}\Big(\lambda\sigma_1- \sum_{j=1}^4 q_j\Big)-
\left(2-{\tfrac  {95} 2}{\tfrac  1 n}+{\tfrac  {30} {n-1}}\right)\left(\sigma_1\sigma_2+\sigma_3\right),\label{r:e}\\
(\sigma_4)^2r_{n-6}&=&{\tfrac \lambda {n(n-1)}}+\left(1-{\tfrac  {33} n}+{\tfrac  {45} 2}{\tfrac  1 {n-1}}\right)\left((\sigma_1)^2+2\sigma_2\right),\label{r:f}\\
(\sigma_4)^2r_{n-7}&=&-\left(2-{\tfrac  {175} 2}{\tfrac  1 n}+{\tfrac  {63} {n-1}}\right)\sigma_1,\label{r:g}\\
(\sigma_4)^2r_{n-8}&=&1-{\tfrac  {56} n}+{\tfrac  {42} {n-1}}.\label{r:h}
\een
\end{subequations}

Using the two initial conditions for the recurrence relation \eqref{frecurrence}:
\begin{subequations}\label{inicond:1,2}
\ben
f_{-7}=0,...f_{-1}=0,\,f_0=1,\, f_1=0,\la{inicond:1}\\
f_{-7}=0,...f_{-1}=0,\,f_0=0,\, f_1=1,\la{inicond:2}
\een
\end{subequations}
we obtain two linearly independent solutions of Eq. \eqref{dF} $\mathcal{F}_{1,2}(z)$.
Both of them are analytical functions in some vicinity of $z=0$
and define the general solution:  $\mathcal{F}(z)=C_1\mathcal{F}_{1}(z)+C_2\mathcal{F}_{2}(z)$ ($C_{1,2}=\text{const}$),
having standard properties of a fundamental basis of solutions of Eq. \eqref{dF}:
\begin{subequations}\label{F:1,2}
\ben
\mathcal{F}_{1}(0)=1,\quad \mathcal{F}_{2}(0)=0, \la{F:1}\\
\mathcal{F}_{1}^\prime(0)=0,\quad \mathcal{F}_{2}^\prime(0)=1.\la{F:2}
\een
\end{subequations}
In addition, these solutions obey the relation
\ben
\mathcal{F}_{1}(z)\mathcal{F}_{2}^\prime(z)-\mathcal{F}_{2}(z)\mathcal{F}_{1}^\prime(z)=\left(P(0)/P(z)\right)^{1/2}.
\la{WronF1F2}
\een

For example, for any $j=1,2,3,4$ one is able to represent the corresponding general Heun's functions in the novel form
\ben
\text{HeunG}(a_{{}_{G,j}}, \lambda, \alpha_{{}_{G,j}}, \beta_{{}_{G,j}}, \gamma_{{}_{G,j}}, \delta_{{}_{G,j}}, z-z_j)=
\Gamma_j^1\,\mathcal{F}_{1}(z;z_1,...,z_4;q_1,...,q_4;\lambda)+\Gamma_j^2\,\mathcal{F}_{2}(z;z_1,...,z_4;q_1,...,q_4;\lambda)
\la{HGF12}
\een
with some coefficients $\Gamma_{j,1}, \Gamma_{j,2}$ which play a fundamental role in our approach to these functions.
Further detailed study of relation \eqref{HGF12} is outside the scope of the present paper.

\subsection{The symmetric choice of the positions of singular points}

We shall take advantage of the freedom to put the singular points $z_{j=0,...,4}$ in the proper places in the complex plane $\mathbb{C}$
for simplifying, as much as possible, the coefficients \eqref{r:a}-\eqref{r:h} and thus, the very solutions $\mathcal{F}_{1,2}(z)$.
This can be done in a symmetric way by imposing additional conditions on the elementary symmetric functions $\sigma_{j=1,2,3,4}$.
Using the proper Mobius transformation one is able to impose three independent constraints on $z_{j=0,...,4}$
without changing the problem, see Appendix \ref{ApA}.
Obvious simple choice is to reduce the quartic equation $P(z)=0$ to the following biquadratic one: $z^4- 2\cos(2\phi)z^2+1=0$
with the roots
\ben
z_1=e^{i\phi},\,\,z_2=-e^{-i\phi},\,\,z_3=-e^{i\phi},\,\,z_4=e^{-i\phi},
\la{z_j}
\een
by imposing three symmetric constraints
\ben
\sigma_1=\sigma_3=0,\,\, \sigma_4=1,
\la{B_sigma_constraits}
\een
and replacing $\sigma_2= -2\cos(2\phi)$ with one more complex uniformization parameter $\phi\in \mathbb{C}$.
This time our goal is to avoid branching points of the roots of the above biquadratic equation.

The meaning of the new variable $\phi$ is revealed by the formula for the invariant $a(z_1,z_2,z_3,z_4)$
of the Mobius transformation -- the so called {\it cross-ratio}, see Appendix \ref{ApA}.
In our problem it acquires the form
\ben
a={\frac {(z_1-z_3)(z_2-z_4)}{(z_2-z_3)(z_1-z_4)}}={\frac 1{\left(\sin\phi\right)^2}}
\quad \Rightarrow\quad\sigma_2=-2\left(1-{\frac 2 a}\right).
\la{a}
\een

Now, it is not hard to obtain the relations
\begin{subequations}\label{rho:2,3,4,5}
\ben
\sum_{k=1}^4 q_j/z_j &=& +{\tfrac i 4}\,\sin(2\phi)\,\rho_2,\label{rho:2}\\
\sum_{k=1}^4 q_j\sigma_2^j &=&-{\tfrac i 4}\,\sin(2\phi)\,\rho_3,\label{rho:3}\\
\sum_{k=1}^4 q_j\sigma_1^j &=&-{\tfrac i 4}\,\sin(2\phi)\,\rho_4,\label{rho:4}\\
\sum_{k=1}^4 q_j &=&+{\tfrac i 4}\,\sin(2\phi)\,\rho_5, \label{rho:5}
\een
\end{subequations}
where we introduce the following four elementary functions of five variables $\{\phi,\chi_1,\chi_2\chi_3,\chi_4\}$:
\begin{subequations}\label{rh:2,3,4,5}
\ben
\rho_2=\hskip .8truecm\left(\left(\sin(2\chi_1)\right)^2+\left(\sin(2\chi_3)\right)^2\right)&-&\hskip .7truecm\left(\left(\sin(2\chi_2)\right)^2+\left(\sin(2\chi_4)\right)^2\right),\\
\rho_3=e^{-i\phi}\left(\left(\sin(2\chi_1)\right)^2-\left(\sin(2\chi_3)\right)^2\right)&+&\,
e^{i\phi}\,\left(\left(\sin(2\chi_2)\right)^2-\left(\sin(2\chi_4)\right)^2\right),\\
\rho_4=e^{2i\phi}\,\left(\left(\sin(2\chi_1)\right)^2+\left(\sin(2\chi_3)\right)^2\right)&-&
e^{-2i\phi}\left(\left(\sin(2\chi_2)\right)^2+\left(\sin(2\chi_4)\right)^2\right),\\
\rho_5=\,e^{i\phi}\,\,\left(\left(\sin(2\chi_1)\right)^2-\left(\sin(2\chi_3)\right)^2\right)&+&\,
e^{-i\phi}\left(\left(\sin(2\chi_2)\right)^2-\left(\sin(2\chi_4)\right)^2\right).
\een
\end{subequations}

As a result, one obtains much simpler formulas for the coefficients in recurrence \eqref{frecurrence}:
\begin{subequations}\label{r134:abcdefgh}
\ben
\hskip -.6truecm r_{n-1}&=&0,\label{r134:a}\\
\hskip -.6truecm r_{n-2}&=&{\tfrac 1 {n(n-1)}}\Big(\lambda- {\tfrac i 4}\,\sin(2\phi)\,\rho_2\Big)+
4\left(1-{\tfrac  5 n}+{\tfrac  3 2}{\tfrac  1 {n-1}}\right)\cos(2\phi), \label{r134:b}\\
\hskip -.6truecm r_{n-3}&=&-{\tfrac 1 {n(n-1)}}{\tfrac i 4}\,\sin(2\phi)\,\rho_3, \label{r134:c}\\
\hskip -.6truecm r_{n-4}&=&{\tfrac 1 {n(n-1)}}\Big(\!-2\lambda \cos(2\phi)+ {\tfrac i 4}\,\sin(2\phi)\,\rho_4\Big)+
2\left(1-{\tfrac  {16} n}+{\tfrac  9 {n-1}}\right)\left((\cos(2\phi)^2+1\right),\label{r134:d}\\
\hskip -.6truecm r_{n-5}&=&{\tfrac 1 {n(n-1)}}{\tfrac i 4}\,\sin(2\phi)\,\rho_5, \label{r134:e}\\
\hskip -.6truecm r_{n-6}&=&{\tfrac \lambda {n(n-1)}}-4\left(1-{\tfrac  {33} n}+{\tfrac  {45} 2}{\tfrac  1 {n-1}}\right)\cos(2\phi),\label{r134:f}\\
\hskip -.6truecm r_{n-7}&=&0,\label{r134:g}\\
\hskip -.6truecm r_{n-8}&=&1-{\tfrac  {56} n}+{\tfrac  {42} {n-1}}.\label{r134:h}
\een
\end{subequations}

If in addition to constraints \eqref{B_sigma_constraits} one imposes one more symmetric constraint, namely:
\ben
\sigma_2=0,
\la{sigma_2_0}
\een
then $a=2$, $\phi / \hskip -.2truecm \mod\!(2\pi)=\pm\pi/4,\pm 3\pi/4$,
and one obtains the simplest possible coefficients in recurrence \eqref{frecurrence}:
\begin{subequations}\label{sr134:abcdefgh}
\ben
r_{n-1}&=&0,\label{sr134:a}\\
r_{n-2}&=&{\tfrac \lambda {n(n-1)}}, \label{sr134:b}\\
r_{n-3}&=&0, \label{sr134:c}\\
r_{n-4}&=&2\left(1-{\tfrac  {16} n}+{\tfrac  9 {n-1}}\right),\label{sr134:d}\\
r_{n-5}&=&0, \label{sr134:e}\\
r_{n-6}&=&{\tfrac \lambda {n(n-1)}},\label{sr134:f}\\
r_{n-7}&=&0,\label{sr134:g}\\
r_{n-8}&=&1-{\tfrac  {56} n}+{\tfrac  {42} {n-1}}.\label{sr134:h}
\een
\end{subequations}

Note that the additional constraint \eqref{sigma_2_0} brings us not only to a simplification
of the coefficients in recurrence \eqref{frecurrence},
but also restricts the class of the solutions of Eq. \eqref{dF} under consideration.

\section{The circular case}

The case of the general Heun's functions when the singular points of Eq. \eqref{dHeunG}
are placed on the real axis $\mathbb{R}$ by construction \cite{Smirnov} is circular one,
since $\mathbb{R}$ can be considered as a circle with an infinite radius.
The Mobius transformation preserves the circular property, see Appendix \ref{ApA},
as well as \cite{Smirnov}, where the circular case for the general Heun's equation was substantially
elaborated using the standard Eq. \eqref{dHeunG} and without any relation with the choice \eqref{z_j} in Eq. \eqref{dF}.

The value of the invariant cross-ratio \eqref{a} $a\in\mathbb{R}$ is real
for any four complex points $z_{j=1,2,3,4}$ on a circle in $\mathbb{C}$.
Then, from relation \eqref{a} follows that in the circular case the angle $\phi\in\mathbb{R}$ is real and
the singular points \eqref{z_j} lie on the unit circle with the center $z=0$.
According to the basic results of the standard analytic theory of ordinary differential equations
\cite{Golubev,CL,KF}, we obtain our key result:

{\bf Proposition 5:} In the circular case, the series \eqref{taylorF} with coefficients \eqref{r134:a}-\eqref{r134:h}
and $\phi\in\mathbb{R}$ are absolutely convergent inside the unit circle, i.e., for any $z\in \mathbb{C}$ with $|z|<1$.\,$\blacktriangleleft$

{\bf Corollary:} In the circular case, the four regular singular points $z_{j=1,2,3,4}$
of the general Heun's equation \eqref{dF} can be treated equally from inside the unit circle
using the Taylor series \eqref{taylorF}.

Next important step is to restrict Proposition 4 (iii) to the circular case.

{\bf Proposition 6:}
In the circular case, equation  \eqref{dF} preserves its form if one makes the following substitutions
\ben
z\to 1/z,\quad, z_j \to 1/z_j, \quad F(z)\to F(1/z)\quad P(z)\to P(1/z),\nonumber \\
\lambda \to \lambda -{\tfrac i 4}\sin(2\phi)\rho_2,\quad q_j \to -q_j/z_j^2.
\la{inverssion_circular}
\een
This way we obtain from solutions \eqref{taylorF} new solutions or Eq. \eqref{dF} in the form of the
Laurent series expansions which are absolutely convergent for any $z\in \mathbb{C}$ with $|z|>1$.\,$\blacktriangleleft$

{\bf Corollary:} In the circular case, the four regular singular points $z_{j=1,2,3,4}$
of the general Heun's equation \eqref{dF} are mapped under inversion on the same points,
removed to the initial positions $z_{j=4,3,2,1}$, respectively. Hence,
one can treat all singular points equally from outside the unit circle
using the corresponding  Laurent series, described in Proposition 6.

As a final result, in the circular case we reach a totally symmetric treatment of the singular points
$z_{j=1,2,3,4}$  in the whole complex plane $\mathbb{\tilde C}$.

\section{Some comments and concluding remarks}
In the present paper, we introduced and studied a novel representation of the general Heun's functions.
It is based on the symmetric form of the Heun's differential equation
yielded by a further development of the Felix Klein
symmetric form of the Fuchsian equations with an
arbitrary number $N\geq 4$ of regular singular points.
We derived the symmetry group of these equations and their solutions.
It turns to be a proper extension of the Mobius group.

The basic relations for the general Heun's equation with $N=4$ are derived and discussed in detail.

Special attention was paid to the nine-term recurrence relation for the coefficients of the Taylor series solutions
of the novel symmetric form of the general Heun's equation.
We described in detail the simplification of these coefficients
using the proper Mobius transformation of the singular points.

We also showed that in the circular case, 
when the four singular points of the symmetric form of the gemeral Heun's equation lie on the unite circle, 
the novel Taylor series solutions are absolutely convergent inside it.
After the simple inversion of the independent variable $z\to 1/z$ one obtains also  the Laurent series solutions
which are absolutely convergent outside the circle with unit radius.
Hence, in the circular case one can use the new solutions for a simultaneous equal treatment of all singular points.

A more detailed study of the basic relation \eqref{HGF12},
as well as consideration of the corresponding confluent cases
of the Heun's equation will be presented elsewhere.

One can hope that this new approach will simplify
the solution of the existing basic open problems in the theory of the general Heun's functions.
The novel representation will allow also development of new effective computational methods
for calculations with the Heun's functions
which at present are still a quite problematic issue.

\vskip .7truecm
\noindent{\bf \Large Acknowledgments}
\vskip .3truecm
The author is grateful to Professor Sergei Yu. Slavyanov for his helpful comments,
help with the literature, and his kind encouragement,
to Professors Alexander Kazakov and Artur Ishkhanyan, as well as
to other participants in the Section {\it Heun's equations and their applications}
of the Conference {\it Days on Diffraction 2014}, St. Petersburg, May 26-30, 2014
for their interest in the talk on the basic results of the present article
which were reported there for the first time, 
as well as to Professor Oleg Motygin for his useful remarks.

Special thanks are also to the leader of the Maple-Soft-developers, Edgardo Cheb-Terrap,
for many useful discussions during the last years
on the properties of the Maple-Heun's functions and the computational problems with them.

The author is also thankful to the leadership of BLTP, JINR, Dubna for the support and good working conditions.
This article was also supported by the Sofia University Foundation {\it Theoretical and Computational Physics and Astrophysics}
and by the Bulgarian Agency for Nuclear Regulation, the 2014 grant.

\appendix

\section{The basic properties of Mobius transformation}\label{ApA}
In this appendix we present some well-known basic properties of the Mobius transformation just for reader's convenience
(see, for example, \cite{Golubev,CL,KF,Smirnov,AF} for more detail.).

This fractional-linear transformation of the compactified complex plane $\mathbb{\tilde C}$ (the Riemann sphere)
is the one-to-one mapping $\mathbb{\tilde C}\leftrightarrow\mathbb{\tilde C}$ defined by the formulas
\ben
z\leftrightarrow u: \quad u={\frac {az+b}{cz+d}}\leftrightarrow z={\frac {du-b}{-cu+a}},\quad \forall z,u \in \mathbb{\tilde C}, \quad ad-bc\neq 0.
\la{Mobius}
\een
It has the following well-known basic properties which we use in the present paper.

1. Decomposition property:
\ben
{\frac {az+b}{cz+d}}=\left(z+{\frac a c}\right)\circ\left({\frac {bc-ad} {c^2}}z\right)\circ\left({\frac 1 z}\right)\circ\left(z+{\frac d c}\right).
\la{decomp}
\een
In Eq. \eqref{decomp}, a small circle denotes the composition of the basic maps $\mathbb{\tilde C}\leftrightarrow\mathbb{\tilde C}$:
translation: $z\rightarrow z+{\frac d c}$, inversion: $z\rightarrow {\frac 1 z}$,
homothety and rotation:  $z\rightarrow {\frac {bc-ad} {c^2}}z$,  and (second) translation: $z\rightarrow z+{\frac a c}$.

From Eq. \eqref{decomp} follows that the fractional-linear transformations form the specific Mobius group
$\mathfrak{G}_{Mobius}$ with respect to the composition of the one-to-one mappings $\mathbb{\tilde C}\leftrightarrow\mathbb{\tilde C}$.

2. It preserves the value of the cross-ratio, i.e., for any four complex points $z_{1,2,3,4}\leftrightarrow u_{1,2,3,4}$:
\ben
a={\frac  {(z_1-z_3)(z_2-z_4)} {(z_2-z_3)(z_1-z_4)} }= {\frac {(u_1-u_3)(u_2-u_4)}{(u_2-u_3)(u_1-u_4)}}=\text{invariant}.
\la{corosratio}
\een

3. The necessary and sufficient condition for the cross-ratio \eqref{corosratio} to be a real number $a\in\mathbb{R}$
is that the four points $z_{1,2,3,4}\in \mathbb{\tilde C}$ lie on a circle $\mathfrak{C}\in \mathbb{\tilde C}$.

4. The fractional-linear transformation \eqref{Mobius} is the only univalent complex change of the variable $z\in \mathbb{\tilde C}$
which does not change the number and the character of the singular points of any function on
$\mathbb{\tilde C}$. As a result, it does not change the number and the character of the singular points
of any analytical ordinary differential equation  on $\mathbb{\tilde C}$.



\end{document}